\documentclass[aps,prl,twocolumn,floatfix,showpacs,showkeys,byrevtex]{revtex4}
\usepackage{graphicx}
\usepackage{color}
\usepackage{dcolumn}

\def\sfrac#1#2{{\textstyle{{#1}\over {#2}}}}  

\newcommand{\expup}[1]{e^{#1}}

\newcommand{\EG}{{\textrm{e.g.}}}
\newcommand{\IE}{{\textrm{i.e.}}}
\newcommand{\EA}{{\textit{et al.}}}

\date{\today}

\begin{document}
\title{How low-energy weak reactions can constrain
  three-nucleon forces and the neutron-neutron scattering length}

\author{A. G{\aa}rdestig}\email{anders@physics.sc.edu}
\altaffiliation[Present address: ]{ Department of Physics and Astronomy,
University of South Carolina, Columbia, SC 29208}
\author{D. R. Phillips}\email{phillips@phy.ohiou.edu}
\affiliation{Department of Physics and Astronomy, 
Ohio University, Athens, OH 45701}

\begin{abstract}
We show that chiral symmetry and gauge invariance enforce relations
between the short-distance physics that occurs in a number of
electroweak and pionic reactions on light nuclei.  Within chiral
perturbation theory this is manifested via the
appearance of the same axial isovector two-body contact term in
$\pi^-d\to nn\gamma$, $p$-wave pion production in $NN$ collisions,
tritium $\beta$ decay, $pp$ fusion, $\nu d$ scattering, and the hep
reaction.  Using a Gamow-Teller matrix element obtained from
calculations of $pp$ fusion as input we compute the neutron spectrum
obtained in $\pi^- d \to nn\gamma$.  With the short-distance physics
in this process controlled from $pp\to de^+\nu_e$ the theoretical
uncertainty in the $nn$ scattering length extracted from $\pi^-d\to
nn\gamma$ is reduced by a factor larger than three, to $\alt0.05$~fm.
\end{abstract}



\maketitle

In Quantum Chromodynamics (QCD) the up and down quark masses are much
smaller than the typical scale of hadron masses, $M_{\rm QCD} \sim 1$
GeV/$c^2$, and so the QCD Lagrangian has an approximate $SU(2)_{\rm
  L}\times SU(2)_{\rm R}$ symmetry.  This symmetry is spontaneously
broken by the QCD vacuum, leaving only $SU(2)_{\rm isospin}$ as a
manifest symmetry of low-energy hadronic processes. The
spontaneously-broken symmetry results in the existence of three
(pseudo-)Nambu-Goldstone bosons, the pions, and means that the
quantum-field-theoretic operator corresponding to the Noether currents
of the broken generators changes pion number by one.
External electroweak fields couple to these currents, and so the
chiral $SU(2)_{\rm L}\times SU(2)_{\rm R}$ symmetry of QCD imposes
relationships between the pion-baryon couplings and the axial
couplings of baryons measured in electroweak processes.  One of these
is the Goldberger-Treiman relation:
\begin{equation}
\frac{g_{\rm A}}{f_\pi}=\frac{g_{\pi NN}}{M},
\label{eq:GT}
\end{equation}
which expresses the $\pi NN$ coupling constant $g_{\pi NN}$ in terms
of the nucleon's axial coupling $g_{\rm A}$ and the pion decay constant
$f_\pi$. ($M$ is the nucleon mass.)

The symmetries of QCD have other consequences for
low-energy reactions involving pions. Introducing a
photon field into the theory according to the dictates of
$U(1)_{{\rm em}}$ gauge invariance shows that the leading piece of
the amplitude for the photoproduction of $\pi^+$ or $\pi^-$ at threshold
is given by the Kroll-Ruderman (KR) term:
\begin{equation}
|{\cal A}_{\rm KR}|=\frac{e g_{\rm A}}{f_\pi}.
\label{eq:KR}
\end{equation}

In the single-nucleon sector the combination of $U(1)_{\rm em}$ and QCD's
spontaneously-broken $SU(2)_{\rm L} \times SU(2)_{\rm R}$ enforces relations 
like Eqs.~(\ref{eq:GT}) and (\ref{eq:KR}).  In this paper we explore the
analog of these relations in the two-nucleon sector.  We first exhibit
a relationship between the short-distance physics in $NN \rightarrow
NN \pi$ reactions and electroweak processes such as $pp$ fusion,
tritium beta decay, and muon absorption on deuterium $\mu^-d\to nn\nu_\mu$. 
We then show that the same two-nucleon physics occurs
in some reactions involving photons, in particular in the process $\pi^-
d\to nn\gamma$. The two-nucleon coupling in this reaction is
then directly proportional to the two-nucleon coupling of the axial
current.  Knowledge of one coupling therefore yields the other, so that
precise information about $\pi^- d \rightarrow nn \gamma$ (or its
crossed partner $\gamma d\to nn \pi^+$) constrains
$pp$ fusion.  Conversely, calculations of tritium beta decay and
$pp$ fusion contribute to model-independent predictions for $\pi^-
d\to nn\gamma$.

Equations (\ref{eq:GT}) and (\ref{eq:KR}) are exact only in the chiral
limit of QCD, \IE, when $m_u=m_d=0$. In our world they are satisfied
to within 2\% and 6\% respectively, since they have corrections
proportional to $m_u$ and $m_d$. The tool for
systematically computing such corrections is chiral perturbation
theory ($\chi$PT) (see Ref.~\cite{ulfreview} for a review).  $\chi$PT
is an effective field theory (EFT) in which amplitudes are obtained in
an expansion in $\chi\equiv\frac{m_\pi}{M_{\rm QCD}}$.  Since its
initial development~\cite{We79,GL85} $\chi$PT has been successfully
applied to a number of processes in the meson-meson and meson-nucleon
systems.  But few-nucleon systems involve states which are bound by
the strong interaction, and so the $\chi$PT expansion must be
modified.  Weinberg~\cite{We90} has pointed out that the $\chi$PT
expansion can be applied to the operator $\hat{O}$ used in the
computation of a matrix element
\begin{equation}
  {\cal M}=\langle \psi_f|\hat{O}|\psi_i \rangle
\label{eq:M}
\end{equation}
for reactions involving such systems.
The result---provided consistently-computed wave functions 
$|\psi_{i,f} \rangle$~\cite{bira,evgeny} are used---is an expansion of
${\cal M}$ in powers of $\chi$ that incorporates the
model-independent consequences of the (broken) chiral symmetry
of QCD into calculations of processes in few-nucleon systems.

The operator $\hat{O}$ in (\ref{eq:M}) is computed up to a given
order in the $\chi$PT expansion in powers of $\chi$, using the Feynman
rules derived from the $\chi$PT Lagrangian. At leading-order the
relevant terms for one nucleon are:
\begin{equation}
{\cal L}_N = N^\dagger [i v \cdot D + g_{\rm A} S \cdot u] N
\label{eq:LN}
\end{equation}
where $v$ is the nucleon four-velocity, $D$ is a (chiral) covariant derivative,
$S$ is the Pauli-Lubanski spin vector, and $g_{\rm A}$ parameterizes the 
unknown short-distance physics of the nucleon's axial-current matrix element. 
Here $u_\mu$ is an axial four-vector which contains the pion field, and when 
expanded takes the form:
\begin{equation}
f_\pi u_\mu = -\tau_a \partial_\mu \pi_a - \epsilon_{3 ab}
V_\mu \pi_a \tau_b + f_\pi A_\mu + \mathcal{O}(\pi^3)
\label{eq:umu}
\end{equation}
where $V_\mu$ ($A_\mu$) is an external vector (axial) field.
The Goldberger-Treiman and KR relations (\ref{eq:GT}) and (\ref{eq:KR}) follow 
directly from inserting (\ref{eq:umu}) into the Lagrangian (\ref{eq:LN}) and 
examining the relevant $\pi NN$ and $\gamma\pi NN$ couplings.

The axial four-vector $u_\mu$ also appears when pions, photons, and
axial fields couple to pairs of nucleons. The leading effects of this
can be represented by one term in the chiral
Lagrangian~\cite{Parkhep,HvKM}:
\begin{eqnarray}
  \mathcal{L}_{NN} & = & -2 d_1  
  N^\dagger S \cdot u N N^\dagger N.
\label{eq:LNN}
\end{eqnarray}
In principle the low-energy constant (LEC) $d_1$ can be evaluated by
lattice methods~\cite{DS03}. As with most single-nucleon sector LECs,
however, its value has at present only been obtained from experimental
data. Once the value of the LEC $d_1$ has been extracted from one
process we can use the result to predict other observables on which it
has an impact.  Equation (\ref{eq:LNN}) thus encodes model-independent
correlations between different reactions which share the same
short-distance physics because of the symmetries of the
chiral Lagrangian.  For the reactions discussed here the
short-distance $NN$ physics is associated with
${}^3S_1\leftrightarrow{}^1S_0$ transitions.  It is convenient to
parameterize it by a dimensionless constant, $\hat{d}$, which is
obtained from $d_1$, and from which the scale $Mf_\pi^2$ has been
removed~\cite{Parkhep,Ga06}.  In two-nucleon systems $\hat{d}$ plays a
role analogous to that of $g_{\rm A}$ in the single-nucleon sector.

Model-independent correlations involving $\hat{d}$ have been used to
predict the rates of reactions which are important for the production
and detection of solar neutrinos. In Ref.~\cite{Parkhep} the constant
$\hat{d}$ was fixed using the well-measured tritium beta-decay
half-life. This allowed the authors of that paper to predict the rate
of two reactions which are important members of the chain that
produce energy in main-sequence stars and result in the emission of
solar neutrinos: the proton-fusion process $pp\to de^+\nu_e$ and the
hep process $^3{\rm He} \, p\to {}^4{\rm He}\, e^+\,\nu_e$.
Importantly, $\hat{d}$ also parameterizes the short-distance physics
in the $\nu(\bar\nu)d$ breakup reactions~\cite{nud} that facilitate
the flavor decomposition of the solar-neutrino signal from data
obtained at the Sudbury Neutrino Observatory (SNO)~\cite{SNO}.  Since
the $pp$ fusion, hep, and $\nu(\bar{\nu}) d$ breakup reactions occur
at too low a rate to be reliably measured in the laboratory, the
correlations exhibited in Refs.~\cite{Parkhep,nud} provide important,
model-independent information on them.  (This connection has also been
explored in potential model calculations~\cite{GTpot} and pionless
EFT~\cite{BCpp}.)  The same Gamow-Teller ${}^3S_1\rightarrow{}^1S_0$
transition occurs in $\mu^-d\to nn\nu_\mu$, and the LEC $\hat{d}$
plays a role there too~\cite{mud}.

If we now insert the expression (\ref{eq:umu}) into the Lagrangian
(\ref{eq:LNN}) we see that if a pion is produced via the reaction $NN
\rightarrow NN \pi$, and is in a $p$-wave relative to the $NN$ system, then the
LEC $\hat{d}$ parameterizes the short-distance physics of the
process. This is the two-body analog of the Goldberger-Treiman
relation (\ref{eq:GT}), and it establishes an intriguing connection
between the reactions involved in the production and detection of
solar neutrinos and the physics of pion production in $NN$
collisions.

\begin{figure}[ht]
\includegraphics*[width=20mm]{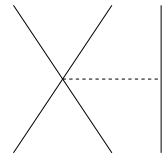}
\caption{The piece of the leading chiral 3NF which involves $\hat{d}$.}
\label{fig:3NF}
\end{figure}

In systems with $A > 2$, $\hat{d}$ plays a subtle, but important, role
in nuclear binding through its generation of a three-nucleon force
(3NF).  As pointed out in Ref.~\cite{3NF}, a 3NF is obtained if a
virtual pion is produced via (\ref{eq:LNN}) and then absorbed on a
third nucleon via the leading Lagrangian (\ref{eq:LN}).  The resulting
diagram is a leading non-vanishing term in the chiral expansion of the
3NF~\cite{3NF,ENGKMW}, and is depicted in Fig.~\ref{fig:3NF}.
Determining $\hat{d}$ accurately from 3N data may be
difficult~\cite{ENGKMW}, but the correlations mandated by the chiral
symmetry of QCD mean that we can use the values of $\hat{d}$ extracted
from electroweak processes---\EG, in Ref.~\cite{Parkhep}---as input to
fix the size of the 3NF depicted in Fig.~\ref{fig:3NF}.  This has the
advantage that the kinematics at which the virtual pion is emitted in
Fig.~\ref{fig:3NF} (pion energy $\approx 0$) matches the kinematics at
which reactions like $pp\to de^+\nu_e$ occur quite well---certainly
much better than it does the kinematics of $NN \rightarrow NN \pi$,
which was used to constrain the 3NF in Ref.~\cite{HvKM}.  This should
reduce the impact that higher-order $\chi$PT corrections have on
predictions for this piece of the 3NF. Qualitative comparison of the
values extracted for $\hat{d}$ in Refs.~\cite{Parkhep} and
\cite{ENGKMW} is encouraging, but for a quantitative comparison
tritium $\beta$-decay and three-nucleon bound state and scattering
calculations will have to be done with consistent wave functions and
regulators.

Here, we are less ambitious.  We will instead exploit another
correlation that results from substitution of (\ref{eq:umu}) into
(\ref{eq:LNN})---one involving photons.  We examine the reaction
$\pi^- d \rightarrow nn\gamma$, which has played a significant role in
determinations of the $nn$ scattering length ($a_{nn}$). Data from
radiative pion capture experiments dominates the accepted value
$a_{nn}=-18.59\pm0.40$~fm~\cite{deTeramond,GGS,LAMPF,NNreview,GP1}.
Final-state interactions in $nd$ breakup experiments can also be used
to measure $a_{nn}$. While one such recent determination is in
agreement with the accepted value, another measurement of $a_{nn}$
using the same reaction disagrees with it
by more than $4\sigma$~\cite{GonzalezTrotterHuhn}.  We
will not comment on this further here, but will focus on
showing that there is a correlation between $\pi^- d \rightarrow nn
\gamma$ and $pp$ fusion. This correlation, together with knowledge of
the $pp$ fusion rate, reduces the impact of short-distance physics on
the value of $a_{nn}$ extracted from $\pi^- d \rightarrow nn \gamma$
data.  This is important because previous extractions of $a_{nn}$ from
this process~\cite{GGS,GP1} have shown that $\pm 0.3$ fm out of the
total uncertainty of $\pm 0.4$ fm in the final result comes from the
short-distance piece of the $NN$ wave function. Up until now it was
not clear how to remove this source of uncertainty, and so reduce the
$a_{nn}$ error below $\pm 0.3$ fm.

We use wave functions that are derived in a manner consistent with chiral
symmetry.  
For the deuteron $S$-state, we start from the asymptotic wave function 
$u^{(0)}(r)=A_S\expup{-\gamma r}$, where $A_S=0.8846$~fm$^{-1/2}$ is the 
asymptotic normalization and $\gamma=\sqrt{MB}=45.7$~MeV the binding 
momentum~\cite{Nijmd}.  
For low-energy $nn$ scattering the starting point is asymptotic wave 
functions with phase shifts derived from the effective-range expansion, \IE,
$v^{(0)}(r)=\sin(pr+\delta)/\sin\delta$, where $p$ is the relative $nn$ c.m.\ 
momentum and $p\cot\delta=-1/a_{nn}+\sfrac{1}{2}r_0p^2$ ($r_0$ is the 
effective range).  
In both cases the wave functions are calculated from $r=\infty$ down
to a matching point $R$ using the one-pion-exchange potential.
For $r<R$ we then assume a spherical well potential and match to the
$r > R$ wave function at $r=R$.  
Further details can be found in Refs.~\cite{GP06,PC,GP1}.
The $pp$ state needed for the calculation of $pp \rightarrow d e^+ \nu_e$ is 
calculated similarly, with the asymptotic wave function given by
zero-energy Coulomb wave functions, and the  
$pp$ scattering length with
respect to the Coulomb potential, 
$a^C_{pp}=-7.8196\pm0.0026$~fm~\cite{app}, as input. 
Higher-order electromagnetic corrections can be included, and give a $\alt1\%$ 
effect~\cite{GTpot}, but for our
present purposes they are not important, since they alter only the
long-distance properties of the $pp$ wave function.

\begin{figure}
\includegraphics*[width=85mm]{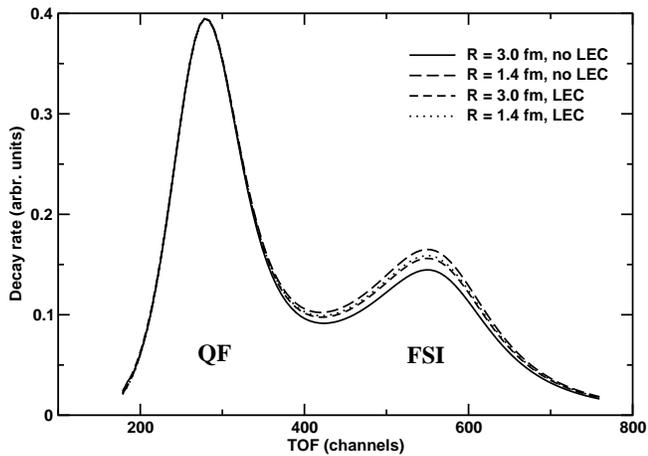}
\caption{Calculated $\pi^-d\to nn\gamma$ neutron time-of-flight distribution, 
for two $R$ values, without and with the LEC $\hat{d}$ contribution.
The supplement of the $\gamma n$ lab.\ angle is 
$\theta_3=0.075$~rad and $a_{nn}=-18.5$~fm.
The labels QF (quasi-free) and FSI indicate where the corresponding 
kinematics are dominant.}
\label{fig:pidR}
\end{figure}

The matrix element for proton fusion is given by~\cite{Parkpp}:
\begin{equation}
  \mathcal{M}_{\rm GT,1} = \frac{1}{A_S}\int dr \, u(r)v^{\rm C}(r),
\label{eq:MGT}
\end{equation}
up to corrections of relative order $\chi^2$. In Eq.~(\ref{eq:MGT})
$u(r)$ and $v^{\rm C}(r)$ are the deuteron $S$-state and $pp$ $^1S_0$
zero-energy scattering (including Coulomb) wave functions.  The matrix
element for the KR contribution to $\pi^-d\to nn\gamma$
in the final-state interaction (FSI) region (see Fig.~\ref{fig:pidR}) is
\begin{equation}
  \mathcal{M}_{\rm FSI} = C \int dr \,
  u(r) \, j_0\left(\frac{kr}{2}\right) \, v(r), 
\label{eq:MFSI}
\end{equation}
where $C$ is a (known) constant, $k$ the photon c.m.\ momentum, 
$j_0$ a spherical Bessel function, and
$v(r)$ the full (energy-dependent) $nn$ $^1S_0$ scattering wave
function in the notation of~\cite{GP1}. Equation (\ref{eq:MFSI}) is also of
accuracy $\chi^2$. 

\begin{figure}[ht]
\includegraphics*[width=85mm]{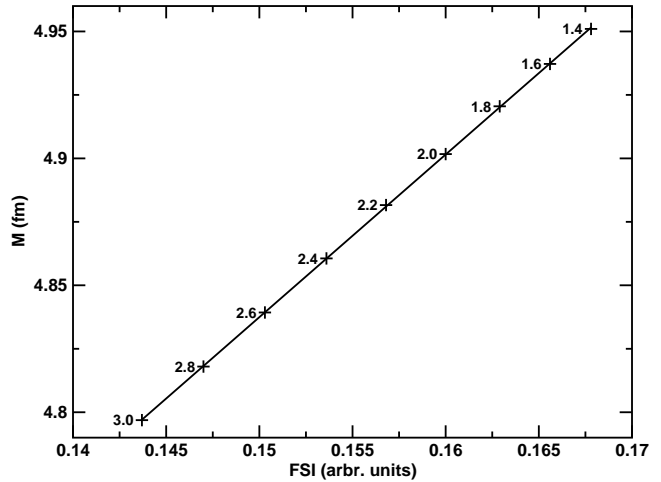}
\caption{Gamow-Teller matrix element~(\ref{eq:MGT}) plotted against 
  $\pi^-d\to nn\gamma$ FSI peak height given by~(\ref{eq:MFSI}) for various 
  values of $R$.  
  The nine points correspond to the values of $R$ (in fm) indicated. 
  The straight line is a linear fit to the points.}
\label{fig:GTvsFSI}
\end{figure}

Figure \ref{fig:GTvsFSI} then shows that the Gamow-Teller matrix
element (\ref{eq:MGT}) is correlated with the FSI peak height in
$\pi^-d\to nn\gamma$, which here, and in 
Fig.~\ref{fig:pidR} below, we have computed using the
$\mathcal{O}(\chi^3)$ $\chi$PT amplitude as in~\cite{GP1}.
The linear correlation in Fig.~\ref{fig:GTvsFSI} suggests that an
essentially $R$-independent result for the height of the FSI peak will
be found if the LEC $\hat{d}$ is adjusted so as to realize a
particular
value of the Gamow-Teller matrix element in $pp$ fusion.

This is implemented in $\chi$PT by adding to the matrix
element (\ref{eq:MGT}) the contribution from $\hat{d}$:
\begin{equation}
  \mathcal{M}^{\rm LEC}_{\rm GT} 
   = \frac{1}{A_S} \hat{d}   \, A_{3S1}\kappa_{3S1} \, A^C_{1S0}\kappa^C_{1S0}.
\end{equation}
Here the short-distance ($r<R$) pieces of the deuteron and $pp$ radial
wave functions are given by $A_{3S1}\sin(\kappa_{3S1}r)$ and
$A^C_{1S0}\sin(\kappa^C_{1S0}r)$.  We then extract a value for
$\hat{d}$ by demanding that the total $\mathcal{M}_{\rm GT}$ agrees
with the result $\mathcal{M}_{\rm GT}=4.898$~fm obtained from the
Argonne $v_{18}$ potential~\cite{AV18} in Ref.~\cite{GTpot,GP06}. (If
we choose the $\mathcal{M}_{\rm GT}$ found using
other $NN$ potentials in Ref.~\cite{GTpot} the change in 
$a_{nn}$ is $< 0.01$~fm.)
This procedure gives the values for
$\hat{d}(R)$ displayed in Table~\ref{tab:CT}.  These values are
``natural''~\cite{Fr96}.

Now $\hat{d}$ contributes to the FSI matrix element through
\begin{equation}
  \mathcal{M}^{\rm LEC}_{\rm FSI} = C\frac{1}{A_S} \hat{d} \,
  A_{3S1}\kappa_{3S1} \, A_{1S0}\kappa_{1S0},
\label{eq:FSId1}
\end{equation}
where $A_{1S0} \sin(\kappa_{1S0} r)$ is the $nn$ (radial) wave function for 
$r<R$.
The $\pi^-d\to nn\gamma$ spectrum (both without and with the $\hat{d}$ 
contribution) for varying radii $R$ is then plotted in
Fig.~\ref{fig:pidR}~\footnote{Only the curve's shape is fitted in
extracting $a_{nn}$, so the curves are slightly rescaled to coincide at the 
QF peak.}. 

Including $\hat{d}$ drastically reduces the dependence on the short-range
physics that dominated the $a_{nn}$ uncertainty in \cite{GGS,GP1}.
If the entire spectrum is fitted our $R$-related 
$a_{nn}$ uncertainty is reduced from $\pm 1.0$~fm to $\pm 0.14$~fm.  Other,
smaller, sources of theoretical error in the $a_{nn}$ extraction are:
the role of $P$-waves in the $nn$ final state, potential changes in
wave functions when chiral two-pion-exchange is incorporated, and
pion-range $\mathcal{O}(Q^4)$ pieces of the two-body operator used to
compute the spectrum of Fig.~\ref{fig:pidR}~\cite{GP1,Ga06}. Now that
it is clear how to remove the $R$-related uncertainty we will include
these effects in our spectrum computation~\cite{GP06}.  It should then
be possible to extract $a_{nn}$ from $\pi^-d\to nn\gamma$ with a
total theoretical uncertainty of $\alt 0.3$~fm if the entire
spectrum is fitted and $\alt0.05$~fm if only the FSI peak is fitted.
In the second case the theoretical error would be more than three
times smaller than in any previous
extraction~\cite{GGS,deTeramond,GP1}.  Experimental uncertainties
would be the dominant contribution to the $a_{nn}$ error bar. Note
that the values found for $a_{nn}$ by fitting in both regions should be
consistent within errors---a useful test of the theory.

\begin{table}
\vspace*{-0.4cm}
\caption{Values of $\hat{d}$ for various matching radii $R$. Here we
have used $f_\pi=93.0$~MeV and $M=939$~MeV.}
\label{tab:CT}
\begin{ruledtabular}
\begin{tabular}{c|cccccc}
  $R$ (fm) & 1.4 & 1.8 & 2.2 & 2.6 & 3.0\\
\hline
$\hat{d}$ & -1.27 & -0.559 & 0.481 & 2.07 & 4.29 \\
\end{tabular}
\end{ruledtabular}
\end{table}

This demonstrates how $\chi$PT links apparently disparate
reactions in few-nucleon systems.  The correlations that result from
this are important not just for their impact on the accuracy with
which the $nn$ scattering length is known.  They also mean that
measurements of, \EG, $\gamma d \rightarrow nn \pi^+$~\cite{Le05},
might help determine the strength of the chiral 3NF, and, if done with
sufficient precision, could test the value of $\hat{d}$ extracted
from tritium beta decay in~\cite{Parkhep}.

We appreciate discussions on these subjects with Malcolm Butler,
Kuniharu Kubodera, and Fred Myhrer. This work was
supported by the Institute for Nuclear and Particle Physics at
Ohio University, by DOE grant DE-FG02-93ER40756, and by 
NSF grant PHY-0457014.

\bibliographystyle{apsrev}

\end{document}